\pdfoutput=1
\documentclass[a4paper,11pt]{article}
\usepackage[utf8]{inputenc}

\usepackage[table]{xcolor}
\usepackage{andocap}
\usepackage{graphicx}
\usepackage{psfrag,fancyhdr,epsfig}
\usepackage{hyperref}
\hypersetup{colorlinks,bookmarksopen,bookmarksnumbered,citecolor=magenta,linkcolor=blue,pdfstartview=FitH,urlcolor=blue}

\usepackage{subfig}
\usepackage{amsmath}
\usepackage{array}
\usepackage{tensor}
\usepackage{orcidlink}

\newcommand{\be}{\begin{equation}}
\newcommand{\ee}{\end{equation}}
\newcommand{\bea}{\begin{eqnarray}}
\newcommand{\eea}{\end{eqnarray}}
\newcommand{\bear}{\begin{array}}
\newcommand{\eear}{\end{array}}
\newcommand{\ba}{\begin{eqnarray}}
\newcommand{\ea}{\end{eqnarray}}

\usepackage[most]{tcolorbox}

\tcbset{colback=yellow!10!white, colframe=red!50!black,
        highlight math style= {enhanced, %<- needed for the 'remember' options
            colframe=red,colback=red!10!white,boxsep=0pt}
        }

\usepackage{empheq}
\newtcbox{\mymath}[1][]{%
    nobeforeafter, math upper, tcbox raise base,
    enhanced, colframe=blue!30!black,
    colback=blue!30, boxrule=1pt,
    #1}

\title{\centering Running cosmological constant in light of recent CMB-BAO data}


\author[a]{Antonio Racioppi}
\author[a]{Mari-Ly Vijar}

\emailAdd{antonio.racioppi@kbfi.ee}
\emailAdd{marilyvijar@gmail.com}

\affiliation[a]{Laboratory of High Energy and Computational Physics,
National Institute of Chemical Physics and Biophysics, R{\"a}vala pst.~10, Tallinn, 10143, Estonia}

\abstract{We revisit single-field slow-roll inflation driven by a running cosmological constant, i.e. a cosmological constant receiving a logarithmic radiative correction, in light of the latest CMB-BAO data. The relative radiative correction $\delta$ is treated as a free parameter of the model and allowed to be positive or negative. For positive $\delta$, the predictions interpolate between a logarithmic regime and a linear inflation one. Compatibility with CMB-only data is confined to a narrow stripe near the $2\sigma$ border where $\delta$ is relatively large. On the other hand, the predictions fit easily within the constraints including BAO data in a wide range of parameters. For negative $\delta$, the slow-roll approximation breaks down at small field values and the predictions are excluded by data. However, once the effective mass contribution inside the logarithmic correction is taken into account, the negative $\delta$ potential develops a local maximum and behaves like a hilltop model, achieving compatibility with CMB-only data, while agreement with CMB-BAO data is possible only at $2\sigma$.}

\begin{document}

\maketitle

%==========================================================================
\section{Introduction}
\label{introduction}
%==========================================================================

Cosmic inflation \cite{Starobinsky:1980te,Guth:1980zm,Linde:1981mu,Albrecht:1982wi} is arguably the most compelling solution to the horizon and flatness problems of standard Big Bang cosmology. In its simplest version, inflation is driven by a single scalar field, the inflaton, slowly rolling down a quasi-flat potential, so that its energy density remains quasi-constant. For several years, the combination of Planck and BICEP/\textit{Keck} (BK) data \cite{Planck:2018jri,BICEP:2021xfz} has favored quasi-flat concave potentials. Such a picture has also been confirmed by the CMB data from the Atacama Cosmology Telescope (ACT) \cite{ACT:2025tim} and the South Pole Telescope (SPT) \cite{SPT-3G:2025bzu}. However, when these data are combined with Planck (the so-called SPA data set), BK and the baryon acoustic oscillation (BAO) measurements of DESI \cite{DESI:2025zgx}, the preferred value of the scalar spectral index $n_s$ is pushed upwards \cite{Ferreira:2025lrd,Balkenhol:2025wms}.

Since inflation requires a quasi-flat potential, a natural choice for the inflaton potential is a running cosmological constant \cite{Shapiro:2000dz,Shapiro:2009dh}, i.e. a cosmological constant receiving a Coleman-Weinberg (CW) \cite{Coleman:1973jx} radiative correction. The use of radiative corrections in inflation is not new. The first models of new inflation \cite{Linde:1981mu,Albrecht:1982wi,Shafi:1983bd} were based on a CW potential. Running cosmological constant models, instead, were first studied in the context of supersymmetric hybrid inflation \cite{Copeland:1994vg,Dvali:1994ms,Binetruy:1996xj,Halyo:1996pp} (see also \cite{Senoguz:2003zw,Rehman:2009nq,Martin:2013tda}), where supersymmetry is spontaneously broken by the vacuum energy during inflation. In such a case, the radiative correction is generated by the mass splitting between the bosonic and fermionic components of the supermultiplets. Since bosons and fermions contribute to the CW potential with opposite signs, the sign of the correction depends on which of them dominates. In supersymmetric hybrid inflation the bosons are heavier, therefore the correction is positive.

More recently, CW potentials have been used again for inflationary model building (e.g. \cite{Barenboim:2013wra,Kannike:2014mia,Kannike:2015apa,Rinaldi:2015yoa,Kannike:2015kda,Marzola:2015xbh,Marzola:2016xgb,Kannike:2016jfs,Artymowski:2016dlz,Dimopoulos:2017xox,Racioppi:2017spw,Karam:2017rpw,Racioppi:2018zoy,Kannike:2018zwn,Racioppi:2019jsp,Gialamas:2020snr,Racioppi:2021ynx,Kannike:2023kzt} and refs.~therein). In particular, it has been shown that, once the inflaton is non-minimally coupled to gravity, the predictions of CW inflation converge towards the ones of linear inflation \cite{Rinaldi:2015yoa,Kannike:2015kda,Racioppi:2017spw,Racioppi:2018zoy}. Radiative corrections have been recently brought back to attention as a way to reconcile inflationary models \cite{Gialamas:2025kef,Wolf:2025ecy,Ahmed:2025rrg,Ellis:2025bzi,Yuennan:2025inm,Yin:2025rrs,Han:2025cwk,Ahmed:2025sfm,Ahmed:2026agd,Keus:2025iwa,Afshar:2025ndm,Ben-Dayan:2026bif,Alexandre:2025ixz,Ellis:2026ceb,Ellis:2026cmx} with the CMB-BAO constraints \cite{ACT:2025tim,Ferreira:2025lrd,Balkenhol:2025wms}. Among them, also the running cosmological constant model has been revisited \cite{Rehman:2025fja,Okada:2025nyd,Ahmed:2026gqq}. However, these works considered specific constructions based on supersymmetry breaking and, therefore, a positive radiative correction of limited size. In this work, instead, we study the running cosmological constant model from an effective theory perspective, without assuming any specific construction. Therefore, we do not impose additional constraints on the size of the radiative correction and we investigate the possibility of both a positive and a negative correction.

This article is organized as follows. In section~\ref{sec:model} we introduce the model and derive the relevant slow-roll quantities. In section~\ref{sec:results} we present our numerical predictions for positive and negative radiative correction, including, to the best of our knowledge for the first time, a regularized version of the latter, and we compare them with both the SPA+BK and the SPA+BK+DESI constraints. In section~\ref{sec:conclusions} we summarize our findings. Finally, in Appendix~\ref{appendix} we discuss how the regularized potential can be derived from the one-loop CW potential.

%==========================================================================
\section{Running cosmological constant and inflation}
\label{sec:model}
%==========================================================================
Our starting point is the usual action describing single-field slow-roll inflation:
\be 
S= \int d^4x\sqrt{-g}\left\{\frac{M_P^2}{2} R -\frac{\partial_\mu \phi \, \partial^\mu \phi}{2} - V(\phi) \right\}, 
\label{eq:Sstart} 
\ee
where $M_P$ is the reduced Planck mass and $\phi$ the inflaton field with potential 
\begin{equation}
V(\phi)=\Lambda^4 \left[1 + \delta \ln\!\left(\frac{\phi}{M}\right)\right] \, , \label{eq:V}
\end{equation}
which is nothing but a running cosmological constant, where $\Lambda$ is the tree-level value of such a constant, $\delta$ the relative radiative correction and $M$ the field value where such a correction is zero (i.e. where the cosmological constant is equal to its tree-level value). Since an exact cosmological constant would imply eternal inflation and/or a scale-invariant spectrum, it seems quite natural to use radiative corrections as a means to trigger the end of inflation and generate a quasi-scale-invariant spectrum. Such a potential is well-known\footnote{In the following we present the most relevant aspects of the model from an effective field theory point of view. A UV complete realization of such a scenario is beyond the scope of the present paper and is left for future work. For more details see Appendix~\ref{appendix} or e.g. \cite{Martin:2013tda} and refs.~therein.} and can be easily generated by the CW potential of a theory containing particles with explicit (non-dynamically generated) mass terms. 
In order to perform an inflationary analysis, we introduce the potential slow-roll parameters:
\begin{eqnarray}
\epsilon(\phi) &=& \frac{M_{\mathrm{P}}^2}{2}\left(\frac{V'(\phi)}{V(\phi)}\right)^2 = \frac{M_{\mathrm{P}}^2}{2}\left(\frac{\delta}{\phi\left[1+\delta\ln(\phi/M)\right]}\right)^2 , \label{eq:eps}\\
\eta(\phi)&=&M_{\mathrm{P}}^2\frac{V''(\phi)}{V(\phi)}=-\,\frac{M_{\mathrm{P}}^2\,\delta}{\phi^2\left[1+\delta\ln(\phi/M)\right]} \, , \label{eq:eta}
\end{eqnarray}
where we used the definition of the potential in eq.~\eqref{eq:V}.
The number of $e$-folds between horizon exit and the end of inflation is given by
\begin{equation}
N_{\mathrm{e}}=\frac{1}{M_{\mathrm{P}}^2}\int_{\phi_{\mathrm{end}}}^{\phi_N}\frac{V(\phi)}{V'(\phi)}\,\mathrm{d}\phi = \frac{1}{M_{\mathrm{P}}^2}\left[\frac{\phi^{2}}{2\delta}+\frac{1}{2}\phi^{2}\ln\!\left(\frac{\phi}{M}\right)-\frac{1}{4}\phi^{2}\right]_{\phi_{\mathrm{end}}}^{\phi_N} , \label{eq:Ne}
\end{equation}
where we used eq.~\eqref{eq:V}, $\phi_N$ is the field value at horizon exit and the end of inflation is defined by the condition
\begin{equation}
\epsilon(\phi_{\mathrm{end}})=1 . \label{eq:epsend}
\end{equation}
In the following we consider the standard range $N_e \in [50,60]$.
The tensor-to-scalar ratio and the scalar spectral index are respectively given by
\begin{equation}
r=16\,\epsilon , \label{eq:ns:r} \qquad n_{\mathrm{s}}=1 - 6\epsilon + 2\eta,
\end{equation}
and applying eqs.~\eqref{eq:eps} and \eqref{eq:eta}, they become
\begin{eqnarray}
r&=&8\,M_{\mathrm{P}}^2\,\frac{\delta^2}{\phi_N^{2}\left[1+\delta\ln(\phi_N/M)\right]^2} , \label{eq:r} \\
n_{\mathrm{s}}&=&1 - 3\,M_{\mathrm{P}}^2\,\frac{\delta^2}{\phi_N^{2}\left[1+\delta\ln(\phi_N/M)\right]^2} - 2\,M_{\mathrm{P}}^2\,\frac{\delta}{\phi_N^{2}\left[1+\delta\ln(\phi_N/M)\right]} . \label{eq:ns}
\end{eqnarray}
Finally, the amplitude of the scalar perturbations is
\begin{equation}
A_{\mathrm{s}}=\frac{1}{24\pi^2 M_{\mathrm{P}}^4}\frac{V(\phi_N)}{\epsilon(\phi_N)} = \frac{\Lambda^4\,\phi_N^{2}\left[1+\delta\ln(\phi_N/M)\right]^3}{12\pi^2\,M_{\mathrm{P}}^{6}\,\delta^{2}}, \label{eq:As}
\end{equation}
which is constrained by Planck data \cite{Planck:2018vyg} to be 
\begin{equation}
A_s\simeq 2.1\times 10^{-9} \, . \label{eq:As:exp}
\end{equation}
\begin{figure}[t]
    \centering
    \includegraphics[width=0.65\textwidth]{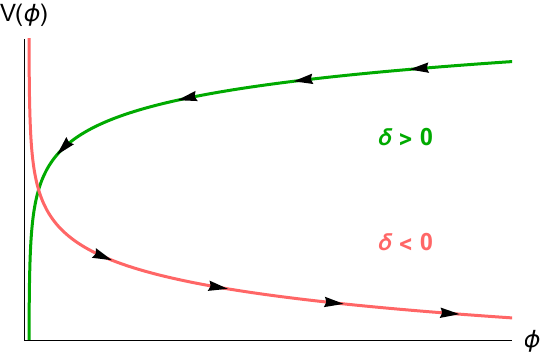}
    \caption{The inflationary potential and the inflaton roll direction for positive (green) and negative (red) values of $\delta$.}
    \label{fig:roll_direction}
\end{figure}
Before proceeding with the numerical analysis, it is worth spending a few more words on the relative radiative correction $\delta$. The parameter $\delta$ can be either positive or negative, according to whether the underlying complete theory has more bosonic or fermionic degrees of freedom dominating the radiative corrections to the cosmological constant \cite{Coleman:1973jx,Shapiro:2009dh}. The sign of $\delta$ has a tremendous impact on the phenomenology as the shape of the potential completely changes, as shown in Fig.~\ref{fig:roll_direction}, where we plot the generic shape of $V(\phi)$ and the inflaton roll direction for positive (green) and negative (red) values of $\delta$. The potential remains positive for $\delta>0$ ($<0$) when $\phi >\phi_0$ ($<\phi_0$), where $\phi_0$ is defined as
\begin{equation}
\phi_0 = M\, e^{-1/\delta} \, . \label{eq:phi:0}
\end{equation}
Therefore, the expression for the potential in eq.~\eqref{eq:V} needs to be interpreted as an effective expression valid for $\phi >\phi_0$ ($<\phi_0$) when $\delta>0$ ($<0$), while outside the aforementioned range higher-order corrections are needed in order to keep the potential positive. As we can see from Fig.~\ref{fig:roll_direction}, for $\delta>0$ ($<0$), the inflaton rolls from larger (smaller) values down to smaller (larger) values in the direction of $\phi_0$. 
Given the fundamental difference between the positive and negative $\delta$ cases, they will be treated separately in the next section.

%==========================================================================
\section{Results}
\label{sec:results}
%==========================================================================

In this section we discuss the phenomenological results for the two choices $\delta>0$ and $\delta<0$. For convenience, we define $\lambda\equiv\Lambda/M_P$.

\subsection{$\delta>0$}
\label{sec:posdelta}
\begin{figure}[t]%
    \centering
    \includegraphics[width=\textwidth]{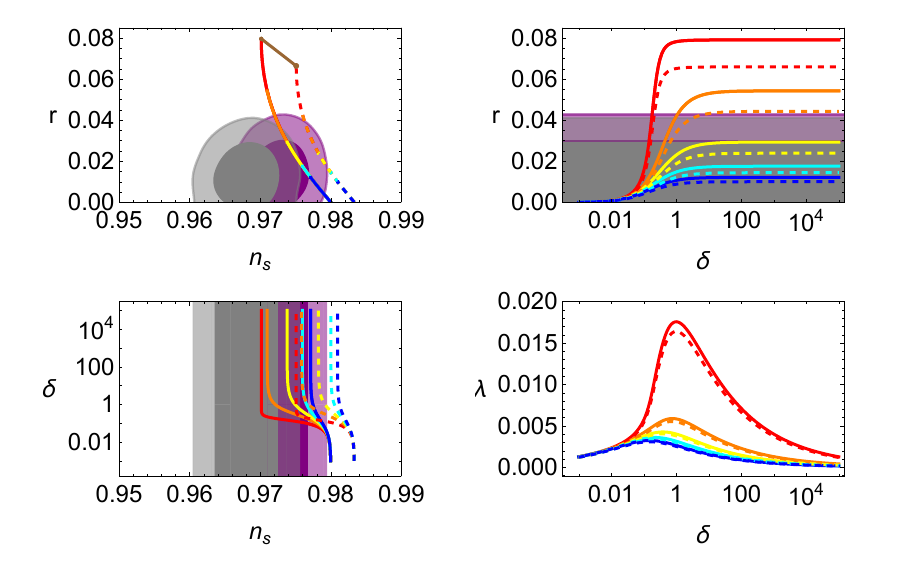}
   \caption{$r$ vs.\ $n_s$ (top left), $r$ vs.\ $\delta$ (top right), $\delta$ vs.\ $n_s$ (bottom left) and $\lambda$ vs.\ $\delta$ (bottom right) with $\delta>0$ and $M/M_{\mathrm P}=10^{-2},\, 10^{-1},\,1,\,10,\,10^{3}$ (blue, cyan, yellow, orange, red) for $N_e=50$ (continuous) and $N_e=60$ (dashed). For reference, we also show the $1\sigma$ and $2\sigma$ constraints from SPA+BK (gray) and SPA+BK+DESI (purple) \cite{Balkenhol:2025wms} and the predictions of linear inflation for $N_e \in [50,60]$ (brown).}
    \label{fig:posdelta}
\end{figure}

In this subsection we present the results for $\delta>0$. 
Fig.~\ref{fig:posdelta} shows $r$ vs.\ $n_s$ (top left), $r$ vs.\ $\delta$ (top right), $\delta$ vs.\ $n_s$ (bottom left) and $\lambda$ vs.\ $\delta$ (bottom right) with $\delta>0$ and $M/M_{\mathrm P}=10^{-2},\, 10^{-1},\,1,\,10,\,10^{3}$ (blue, cyan, yellow, orange, red) for $N_e=50$ (continuous) and $N_e=60$ (dashed). For reference, we show the $1\sigma$ and $2\sigma$ constraints from SPA+BK (gray) and SPA+BK+DESI (purple) \cite{Balkenhol:2025wms}, and the predictions of linear inflation for $N_e \in [50,60]$ (brown).

First of all, we notice that agreement with the SPA+BK constraints is only possible at $N_e \simeq 50$ for $M \gtrsim M_P$ and $\delta \gtrsim 0.1$. On the other hand, agreement with the SPA+BK+DESI data is achieved in a wide range of parameters. For $N_e=50$ a large part of the predictions lies inside the SPA+BK+DESI $1\sigma$ region, while the $N_e=60$ predictions sit, as usual, at slightly higher $n_s$, where they remain compatible at $2\sigma$ with the SPA+BK+DESI data. 
Moreover, at a given $N_e$, the predictions for different values of $M$ overlap and lie on the same line in the $(n_s,r)$ plane. Additionally, the predictions for $r$ and $n_s$ for different values of $M$ share the same asymptotic limit at $\delta \to 0$, but a different one at $\delta \to \infty$, depending on the exact value of $M$. This can be explained as follows. We can write the inflaton potential in terms of $\phi_0$ by using eq.~\eqref{eq:phi:0}, obtaining
\be
 V(\phi)=\delta \,  \Lambda ^4 \ln\left(\frac{\phi }{\phi_0}\right) \, , \label{eq:V:0}
\ee
where we can see that now $\delta$ is only contributing to the normalization of the potential and the inflationary observables 
\bea
N_e &=& \left[\frac{\phi ^2 }{4 M_P^2} \left(2 \ln \left(\frac{\phi }{\phi_0}\right)-1\right)\right]_{\phi_{\mathrm{end}}}^{\phi_N} \, \label{eq:Ne:0}\\
 r &=& \frac{8 M_P^2}{\phi_N ^2 \ln ^2\left(\frac{\phi_N }{\phi_0}\right)}\, , \label{eq:r:0} \\
 n_s &=& 1 -\frac{3 M_P^2}{\phi_N ^2 \ln ^2\left(\frac{\phi_N }{\phi_0}\right)}-\frac{2 M_P^2}{\phi_N ^2 \ln \left(\frac{\phi_N }{\phi_0}\right)}\,  , \label{eq:ns:0}
\eea
depend only on the single parameter $\phi_0$, in agreement with Fig.~\ref{fig:posdelta}. Moreover, even though the parametrization in eq.~\eqref{eq:V} gives a better physical understanding in terms of tree-level and loop contributions, it has the limitation, at a given $M$, of covering only the $\phi_0$ range between 0 and $M$ (see eq.~\eqref{eq:phi:0}). In fact, for $\delta \to \infty$ the predictions approach asymptotic values corresponding to $\phi_0=M$.

Finally, we comment on the two different asymptotic limits, respectively for $\delta \to \infty$ with $M \gg M_P$ and for $\delta \to 0$ (or equivalently $\phi_0 \gg M_P$ and $\phi_0 \to 0$). For $\phi_0 \gg M_P$, the predictions are equivalent to the ones of linear inflation. This is because, in such a case, $\phi_N$ is very close to $\phi_0$; therefore we can perform the change of variable $\varphi = \phi-\phi_0$ (which does not induce a redefinition of the inflaton kinetic term) and expand the potential to first order in $\varphi$, obtaining
\be
 V(\varphi) =   \delta \,  \Lambda ^4 \frac{\varphi}{\phi_0} \, , \label{eq:V:linear} 
\ee
which is a potential linear in $\varphi$. On the other hand, for $\phi_0 \to 0$ we have that $\phi_N \gg \phi_0$. Taking the corresponding asymptotic limits in eqs.~\eqref{eq:Ne:0}-\eqref{eq:ns:0}, we obtain
\bea
r &\approx & \frac{8}{N_e \ln \left(\frac{4 N_e M_P^2}{\phi_0^2}\right)} \to 0 \, , \label{eq:r:log:limit}\\
n_s &\approx & 1-\frac{1}{N_e} -\frac{3}{N_e \ln \left(\frac{4 N_e M_P^2}{\phi_0^2}\right)} \to 1-\frac{1}{N_e} \, , \label{eq:ns:log:limit}
\eea
in agreement with the results of Fig.~\ref{fig:posdelta}. We stress that such a limit is disfavored at more than $2\sigma$ by the SPA+BK+DESI constraints \cite{Balkenhol:2025wms}, even for $N_e=50$.

Regarding the scale of inflation, we can see from the bottom right panel of Fig.~\ref{fig:posdelta} that $\lambda\sim 10^{-3}$-$10^{-2}$, i.e. $\Lambda \sim 10^{15}$-$10^{16}$~GeV, with a maximum around $\delta \sim 1$. Also this behaviour can be easily understood by using eqs.~\eqref{eq:As} and \eqref{eq:V:0}, which give $A_s \propto \delta\Lambda^4 \phi_N^2 \ln^3(\phi_N/\phi_0)$. For small $\delta$, we have $\phi_0 \to 0$, $\ln(\phi_N/\phi_0) \simeq 1/\delta$ and $\phi_N^2 \propto \delta$, so that $A_s \propto \Lambda^4/\delta$ and therefore $\Lambda^4 \propto \delta$ grows with $\delta$. On the other hand, for large $\delta$ the predictions approach their asymptotic values, $A_s$ fixes the combination $\delta\Lambda^4$ and therefore $\Lambda^4 \propto 1/\delta$.

It is worth spending a few words on the range of the parameters $\delta$, $M$ and $\phi_0$. The parameter $\delta$ is a ratio of mass parameters, therefore it is not required by perturbativity to be smaller than one. On the other hand, we need to ensure that the corresponding mass parameter, $\bar\Lambda$, where
\be
 \bar\Lambda^4 = |\delta| \Lambda^4 \, , \label{eq:Lambda:bar}
\ee
is sub-Planckian. It can be easily checked, using eq.~\eqref{eq:V:0}, that the constraint on $A_s$ in eq.~\eqref{eq:As:exp} implies $\bar\Lambda \lesssim 0.03\, M_P$. On the other hand, $M$ and $\phi_0$ can be trans-Planckian, because they essentially correspond to physical masses of the underlying theory (which are sub-Planckian) divided by perturbative (i.e. smaller than one) couplings, with the final effect of a trans-Planckian quantity\footnote{Similar arguments can also be applied for the $\delta<0$ case.}. More details are provided in Appendix~\ref{appendix}. 

Before concluding, we comment on the recent analyses of supersymmetric hybrid inflation in refs.~\cite{Rehman:2025fja,Okada:2025nyd,Ahmed:2026gqq}. In such models the inflationary potential is still approximately given by eq.~\eqref{eq:V} with $M \simeq M_P$, but the radiative correction is fixed by the superpotential and gauge couplings, so that $\delta \ll 1$. 
Therefore, in their minimal version, such models lie in the logarithmic regime discussed above (eqs. \eqref{eq:r:log:limit}-\eqref{eq:ns:log:limit}), which was still compatible at $2\sigma$ with the ACT analysis \cite{ACT:2025fju}, but becomes disfavored once the SPT data are included \cite{SPT-3G:2025bzu,Balkenhol:2025wms}.

To conclude, we stress again that such a scenario fits easily within the SPA+BK+DESI constraints but is strongly disfavored if the constraints do not include the DESI data. Even allowing a relatively large $\delta$, compatibility with the SPA+BK data is confined to a narrow stripe near the $2\sigma$ border.

\subsection{$\delta<0$}
\label{sec:negdelta}

\begin{figure}[t]%
    \centering
    \includegraphics[width=\textwidth]{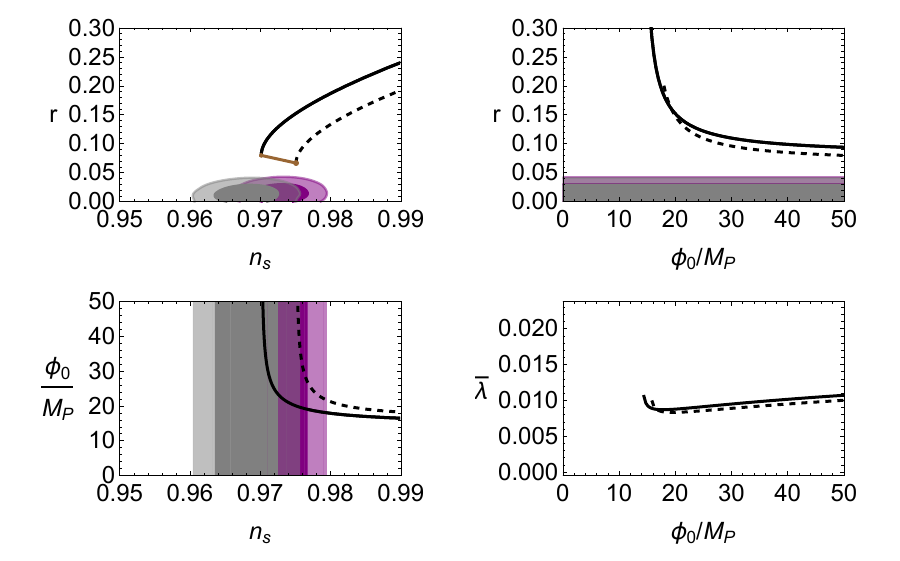}
   \caption{$r$ vs.\ $n_s$ (top left), $r$ vs.\ $\phi_0/M_{\mathrm P}$ (top right),  $\phi_0/M_{\mathrm P}$ vs.\ $n_s$ (bottom left) and $\bar\lambda$ vs.\ $\phi_0/M_{\mathrm P}$ (bottom right) with $\delta<0$ for $N_e=50$ (black, continuous) and $N_e=60$ (black, dashed). For reference, we also show the $1\sigma$ and $2\sigma$ constraints from SPA+BK (gray) and SPA+BK+DESI (purple) \cite{Balkenhol:2025wms} and the predictions of linear inflation for $N_e \in [50,60]$ (brown).}
    \label{fig:negdelta}
\end{figure}

We now study the predictions of the model for $\delta <0$. Since $n_s$ and $r$ depend only on $\phi_0$, we use the parametrization of the potential in eq.~\eqref{eq:V:0}. Moreover, since in this case only $\bar\Lambda$ matters, we define $\bar\lambda = \bar\Lambda/M_P$. 
The numerical predictions are shown in Fig.~\ref{fig:negdelta}, where we present $r$ vs.\ $n_s$ (top left), $r$ vs.\ $\phi_0/M_{\mathrm P}$ (top right),  $\phi_0/M_{\mathrm P}$ vs.\ $n_s$ (bottom left) and $\bar\lambda$ vs.\ $\phi_0/M_{\mathrm P}$ (bottom right) with $\delta<0$ for $N_e=50$ (black, continuous) and $N_e=60$ (black, dashed). For reference, we show the $1\sigma$ and $2\sigma$ constraints from SPA+BK (gray) and SPA+BK+DESI (purple) \cite{Balkenhol:2025wms} and the predictions of linear inflation for $N_e \in [50,60]$ (brown).

We see that there is no value of $\phi_0$ that allows for predictions within the observational constraints, the tensor-to-scalar ratio being always $r\gtrsim0.07$. For large $\phi_0$ the inflaton evolves close to $\phi_0$, where the potential is again approximately linear, and the predictions approach the ones of linear inflation (brown). Lowering $\phi_0$, both $r$ and $n_s$ increase, moving the predictions further away from the allowed regions.

Moreover, $\phi_0$ has a lower bound, sitting around $15\, M_P$ (the exact value depends on $N_e$). The reason is the following. In the case of $\delta<0$, inflation takes place for $0<\phi<\phi_0$, with $\phi_N$ getting closer to 0 as $\phi_0$ decreases. However, for $\phi \to 0$ the first slow-roll parameter in eq.~\eqref{eq:eps} diverges, implying a violation of the slow-roll approximation. This means that, at a given $\phi_0$, the system can produce only a maximum number of $e$-folds of inflation, which decreases with $\phi_0$. The lower bound on $\phi_0$ in the continuous (dashed) line corresponds to this maximum number being exactly 50 (60) $e$-folds. This means that the system is in a regime where eq.~\eqref{eq:V} cannot be trusted. In fact, eq.~\eqref{eq:V} is derived on the assumption that $\phi$ is big enough so that an effective mass term inside the logarithm can be neglected (see Appendix~\ref{appendix}). Clearly, when $\phi_0$ gets too small, such an assumption is no longer valid and such a mass term needs to be considered. We will do that in the next subsection. 

\subsection{Regularized $\delta<0$}
\label{sec:negdelta:reg}

\begin{figure}[t]
    \centering
    \includegraphics[width=0.65\textwidth]{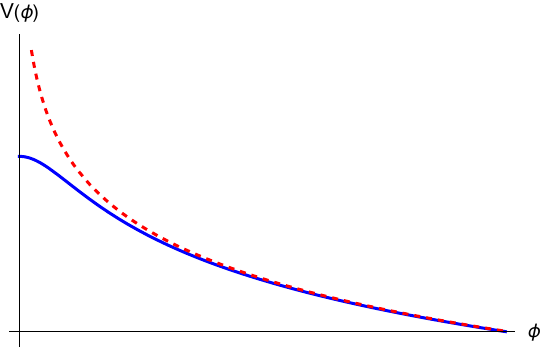}
    \caption{The inflationary potential with $\delta <0$ for the regularized (blue) and unregularized (red, dashed) cases.}
    \label{fig:V:neg:reg}
\end{figure}

\begin{figure}[t]%
    \centering
    \includegraphics[width=\textwidth]{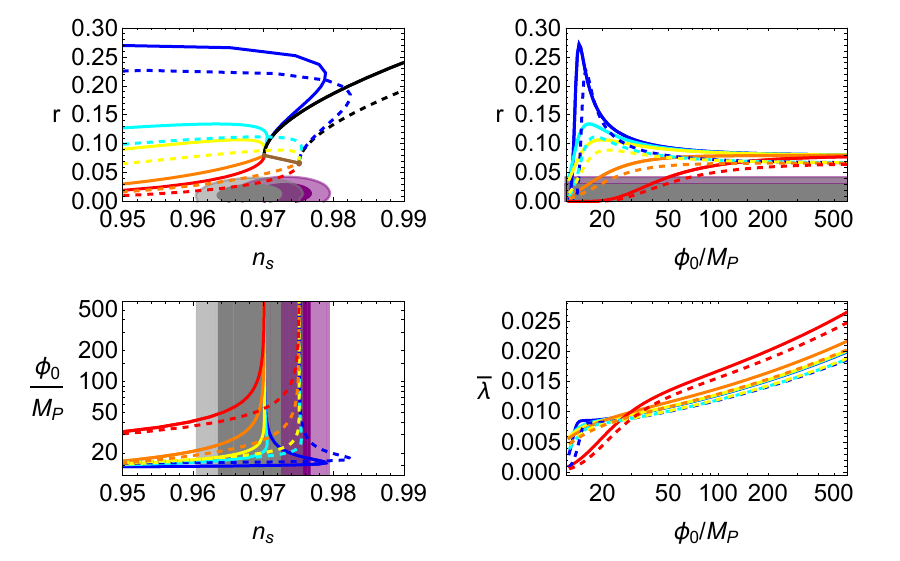}
   \caption{$r$ vs.\ $n_s$ (top left), $r$ vs.\ $\phi_0/M_{\mathrm P}$ (top right),  $\phi_0/M_{\mathrm P}$ vs.\ $n_s$ (bottom left) and $\bar\lambda$ vs.\ $\phi_0/M_{\mathrm P}$ (bottom right) for the potential in eq.~\eqref{eq:V:reg} with $\beta=0.01,\, 0.05,\,0.1,\,0.5,\,0.9$ (blue, cyan, yellow, orange, red) for $N_e=50$ (continuous) and $N_e=60$ (dashed). For reference, we also show the $1\sigma$ and $2\sigma$ constraints from SPA+BK (gray) and SPA+BK+DESI (purple) \cite{Balkenhol:2025wms}, the predictions of linear inflation for $N_e \in [50,60]$ (brown) and the ones of the potential in eq.~\eqref{eq:V:0} (black).}
    \label{fig:negdelta:reg}
\end{figure}

The results of the previous subsection are affected by the unphysical behaviour of eq.~\eqref{eq:V} for $\phi\to 0$. In light of eq.~\eqref{eq:V:0}, the most economical way to take into account the effective mass term inside the logarithm is via\footnote{For more details see Appendix~\ref{appendix}.}
\be
\bar V(\phi) =  -\frac{1}{2} \bar\Lambda^4 \ln \left( \beta +\frac{\phi^2}{\phi_0^2}\right)\, , \label{eq:V:reg}
\ee
where the prefactor $\frac{1}{2}$ is chosen so that, when $\beta \to 0$, the potential in eq.~\eqref{eq:V:0} is restored, with $\bar\Lambda$ given in eq.~\eqref{eq:Lambda:bar}. In Fig.~\ref{fig:V:neg:reg} we show a comparison between a reference negative $\delta$ potential (red, dashed) and the corresponding regularized one (blue), where we can see how at larger $\phi$ values the potentials are essentially indistinguishable, but at lower $\phi$ values one is diverging while the other has a local maximum, hence reproducing a hilltop potential. The local maximum of the potential in eq.~\eqref{eq:V:reg} is located at $\phi=0$ and corresponds to
\be
\bar V(0)=  -\frac{1}{2} \bar\Lambda^4 \ln  \beta  \, . \label{eq:V:max}
\ee
In order to ensure $\bar V(0)>0$, we immediately see that $\beta$ is constrained to be $0<\beta<1$.  The number of $e$-folds for the potential in eq.~\eqref{eq:V:reg} is
\be
 \hspace{-0.28cm} N_e = \left[ \frac{\left(\beta \phi_0^2+\phi^2\right) }{4 M_P^2}  \ln \left(\beta +\frac{\phi ^2}{\phi_0^2}\right) + \beta \frac{\phi_0^2}{4 M_P^2} \left[ \ln \beta  \ln \left(\frac{\phi^2}{\phi_0^2}\right) - \text{Li}_2\left(-\frac{\phi ^2}{\beta  \phi_0^2}\right) \right] - \frac{\phi^2 }{4 M_P^2}\right]^{\phi_N}_{\phi_\text{end}} \!\!\!\! , 
\ee
where $\mathrm{Li}_2$ is the dilogarithm function, and the other inflationary observables are
\bea
 r &=& \frac{32 \phi_N ^2 M_P^2}{\left(\beta  \phi_0^2+\phi_N ^2\right)^2 \ln ^2\left(\beta +\frac{\phi_N ^2}{\phi_0^2}\right)} \, , \\
 n_s &=& 1-\frac{12 \phi_N ^2 M_P^2}{\left(\beta  \phi_0^2+\phi_N ^2\right)^2 \ln ^2\left(\beta +\frac{\phi_N ^2}{\phi_0^2}\right)}-\frac{4 \left(\phi_N
   ^2-\beta  \phi_0^2\right) M_P^2}{\left(\beta  \phi_0^2+\phi_N ^2\right)^2 \ln \left(\beta +\frac{\phi_N ^2}{\phi_0^2}\right)}\, , \\
 A_s &=& -\frac{\bar{\lambda }^4 \left(\beta  \phi_0^2+\phi_N ^2\right)^2 \ln ^3\left(\beta +\frac{\phi_N ^2}{\phi_0^2}\right)}{96 \pi ^2 \phi_N ^2 M_P^2}  \, .
\eea
The corresponding predictions are given in Fig.~\ref{fig:negdelta:reg}, where we show $r$ vs.\ $n_s$ (top left), $r$ vs.\ $\phi_0/M_{\mathrm P}$ (top right),  $\phi_0/M_{\mathrm P}$ vs.\ $n_s$ (bottom left) and $\bar\lambda$ vs.\ $\phi_0/M_{\mathrm P}$ (bottom right) for the potential in eq.~\eqref{eq:V:reg} with $\beta=0.01,\, 0.05,\,0.1,\,0.5,\,0.9$ (blue, cyan, yellow, orange, red) for $N_e=50$ (continuous) and $N_e=60$ (dashed). For reference, we present the $1\sigma$ and $2\sigma$ constraints from SPA+BK (gray) and SPA+BK+DESI (purple) \cite{Balkenhol:2025wms}, the predictions of linear inflation for $N_e \in [50,60]$ (brown) and of the potential in eq.~\eqref{eq:V:0} (black). 
%We recall that $\bar\lambda$ is defined in eq.~\eqref{eq:Lambda:bar}.

As expected, for small $\beta$ (and big $\phi_0$), inflation takes place far away from the regularized region and all curves converge to the unregularized predictions, approaching the linear inflation limit. For small $\beta$, when reducing $\phi_0$, $r$ increases and the predictions depart from the unregularized ones, moving towards very low $n_s$ values. By increasing $\beta$, the predictions depart more and more from the unregularized ones, until they show the typical pattern of hilltop models: very low $n_s$ at very low $\phi_0$, compatibility with the constraints at a relatively high $\phi_0$ and linear inflation limit at very high $\phi_0$. Indeed, near the local maximum, the regularized potential behaves like a hilltop potential (see Fig.~\ref{fig:V:neg:reg}). For $N_e=60$, compatibility with the SPA+BK $2\sigma$ constraints is achievable for\footnote{Even though $\phi_0$ is trans-Planckian and $\beta$ is close to 1, the mass parameters of the underlying theory can be kept sub-Planckian. For more details see Appendix~\ref{appendix}.} $\beta \gtrsim 0.5$, while at $\beta =0.9$ the same is achieved with a bit more than 50 $e$-folds. On the other hand, this last case is compatible with SPA+BK at $1\sigma$ for $N_e=60$. Agreement with the SPA+BK+DESI data is achievable only at $2\sigma$. Regarding the scale of inflation, we can see that again $\bar\lambda\sim 10^{-3}$-$10^{-2}$, i.e. $\bar\Lambda \sim 10^{15}$-$10^{16}$~GeV.

To conclude, the regularized negative $\delta$ scenario shows a behaviour complementary to the positive $\delta$ one: it is in good agreement with the CMB-only constraints, while it is disfavored when the DESI data are included.

%==========================================================================
\section{Conclusions}
\label{sec:conclusions}
%==========================================================================

In this article we have studied single-field slow-roll inflation driven by a running cosmological constant, i.e. a cosmological constant receiving a logarithmic radiative correction, as described by the potential in eq.~\eqref{eq:V}. Differently from previous studies, we have treated the relative radiative correction $\delta$ as a free parameter, considering both of its signs, and we have compared the predictions with the latest SPA+BK and SPA+BK+DESI constraints \cite{Balkenhol:2025wms}.

The sign of $\delta$ completely changes the shape of the potential and the inflaton roll direction, hence the two cases need to be studied separately.
For $\delta>0$, the predictions interpolate between a logarithmic regime, with $n_s \to 1-1/N_e$ and $r\to 0$, and linear inflation. Such a scenario fits easily within the SPA+BK+DESI constraints in a wide range of parameters. On the other hand, compatibility with the SPA+BK data is confined to a narrow stripe near the $2\sigma$ border, which is achievable only for $N_e\simeq 50$ and $\delta \gtrsim 0.1$.
For $\delta<0$, the predictions are excluded by data, the tensor-to-scalar ratio being always $r\gtrsim0.07$. Moreover, the slow-roll approximation breaks down at small field values, imposing $\phi_0 \gtrsim 15\,M_P$. This is a signal that the effective mass term inside the logarithm cannot be neglected anymore. Taking such a term into account, the negative $\delta$ potential develops a local maximum and behaves like a hilltop model. Its predictions are then compatible with the SPA+BK constraints, at $1\sigma$ for $\beta=0.9$ and $N_e=60$, while agreement with the SPA+BK+DESI data is possible only at $2\sigma$.

The two signs of $\delta$ therefore show a complementary behaviour: positive $\delta$ is favored by the SPA+BK+DESI data, while the regularized negative $\delta$ scenario is favored by the CMB-only constraints. A resolution of the CMB-BAO tension \cite{Ferreira:2025lrd,Balkenhol:2025wms} will be crucial in order to discriminate between the two cases. Moreover, since in both cases the predicted tensor-to-scalar ratio is typically $r \sim 10^{-2}$, upcoming experiments, such as the Simons Observatory \cite{SimonsObservatory:2018koc} and LiteBIRD \cite{LiteBIRD:2022cnt}, will provide a further test of such a scenario.

\section*{Note}
This article is partially based on the BSc thesis of M. Vijar \cite{Vijar}.

\acknowledgments
The authors thank C. Dioguardi for useful discussions.
This work was supported by the Estonian Research Council grants TARISTU24-TK10, TARISTU24-TK3, and by the CoE program TK202 ``Foundations of the Universe''. 

\appendix

\section{More on the effective potential} \label{appendix}
The general form of the 1-loop CW potential \cite{Coleman:1973jx} is given by
\be
 V(\phi) = V_\text{tree-level} (\phi) + V_\text{1-loop}(\phi) \, ,
\ee
where the first term is the tree-level contribution
and the CW correction is given by
\be
 V_\text{1-loop}(\phi) = \frac{1}{64\pi^2} \, \mathrm{STr}\left[ \mathcal{M}^4(\phi_i) \left( \ln\frac{\mathcal{M}^2(\phi_i)}{\mu^2} \right) \right] \, , \label{eq:V:CW}
\ee
where $\mathcal{M}^2(\phi_i)$ denotes the field-dependent mass matrices,
$\mu$ is the renormalization scale, and the supertrace
runs over all bosonic and fermionic degrees of freedom, with the latter contributing with a minus sign.
The scheme-dependent constants have been absorbed into a finite redefinition of the tree-level parameters. Since bosons and fermions contribute with opposite signs, the sign of the loop correction can change according to which kind of particles dominates at a given scale. From eq.~\eqref{eq:V:CW} we see that, in a theory with explicit mass terms, the cosmological constant receives a radiative correction. We assume that the dominant part at the inflationary scale is given by a tree-level cosmological constant and the corresponding radiative correction. Focusing then on the inflaton direction (which we label $\phi$), the effective inflationary potential can be parametrized as
\be
 V(\phi) = \Lambda^4 \pm \frac{m_1^4}{64\pi^2}  \ln\left(\frac{m_2^2+y \phi^2}{\mu^2} \right) \, , \label{eq:V:effective:1}
\ee
where $m_{1,2}$ are two effective mass parameters, $y$ an effective coupling and the sign ambiguity depends on whether bosons or fermions dominate the loop correction. We can rewrite the effective potential as
\be
 V(\phi) = \Lambda^4 \left[ 1 + \frac{\delta}{2}  \ln\left(\frac{m_y^2 +\phi^2}{M^2} \right) \right] \, , \label{eq:V:effective:2}
\ee
where
\be
 \delta = \pm \frac{1}{32\pi^2}  \frac{m_1^4}{\Lambda^4} \, , \quad
 M = \frac{\mu}{\sqrt y} \, , \quad m_y = \frac{m_2}{\sqrt y} \, . \label{eq:defs:appendix}
\ee
We can easily see that when $\phi \gg m_y $, eq.~\eqref{eq:V:effective:2} reduces to eq.~\eqref{eq:V}, as expected. Moreover, we see that $\delta$ is not required to be small: being a ratio of mass scales rather than a coupling, $\delta \gtrsim 1$ simply corresponds to a radiative contribution larger than the tree-level cosmological constant, without implying any breakdown of the perturbative expansion, which is controlled by the couplings (e.g. $y$). 

Let us now focus on the $\delta <0$ case.
Before proceeding, it is important to stress that, in such a case, the potential is suitable for inflation only if its local maximum, located at $\phi=0$, is positive, i.e.
\be
 V(0) = \Lambda^4 \left[ 1+ \frac{\delta}{2}  \ln\left(\frac{m_y^2}{M^2} \right) \right] >0 \, . \label{eq:V:bound}
\ee
 Let us now see the relation between eq.~\eqref{eq:V:effective:2} and eq.~\eqref{eq:V:reg}. First of all, we rewrite eq.~\eqref{eq:V:effective:2} as
\be
 V(\phi) =  \Lambda^4 \left[ \frac{\delta}{2}  \ln e^{2/\delta}  + \frac{\delta}{2}  \ln\left(\frac{m_y^2 +\phi^2}{M^2} \right) \right] 
 = \frac{\delta}{2} \Lambda^4   \ln\left(e^{2/\delta} \frac{m_y^2 +\phi^2}{M^2} \right) \, .\label{eq:V:effective:3}
\ee
Defining
\be
 \bar\Lambda^4 = |\delta| \Lambda^4 \, , \quad \beta =  e^{2/\delta} \frac{m_y^2}{M^2} \, , \quad \phi_0=e^{-1/\delta} M \, ,
\ee
we obtain
\be
 V(\phi) =  -\frac{1}{2} \bar\Lambda^4 \ln \left( \beta +\frac{\phi^2}{\phi_0^2}\right) \, ,
\ee
which is exactly the potential given in eq.~\eqref{eq:V:reg}. The constraint $0<\beta<1$ is nothing but a reformulation of the condition given in eq.~\eqref{eq:V:bound}.

Finally, we comment on the trans-Planckian values of $M$ and $\phi_0$ found in our results. Such quantities are field values rather than physical masses and the mass parameters of the underlying theory can always be kept sub-Planckian. First of all, $m_1 = (32\pi^2)^{1/4}\,\bar\Lambda \lesssim 0.1\, M_P$. Moreover, from eq.~\eqref{eq:defs:appendix} we have $m_2 = \sqrt{y}\, m_y$ and $\mu = \sqrt{y}\, M$, while for $\delta<0$ we also have $m_y = \sqrt{\beta}\, \phi_0$. Since the inflationary predictions do not depend on $y$, the choice
\be
 y \lesssim \frac{M_P^2}{\max\left(M^2,\, m_y^2\right)} \label{eq:y:bound}
\ee
ensures sub-Planckian values for both $m_2$ and $\mu$. Such an upper bound is in agreement with perturbativity and with our assumption of neglecting the additional $\phi$-dependent terms induced by the field-dependent masses in eq.~\eqref{eq:V:CW}.

\bibliography{references}

\end{document}